# In Search of a Key Value Store with High Performance and High Availability

Huaibing Jian[†]  Yuean Zhu[†]  Yongchao Long[†]   Bin Li[†]
Shu Wang[†]  Xiliang Wu[†]  Zhichu Zhong[§]

[†]YY Inc.        [§]Jiaying University of China

[†]{jianhuaibing1, zhuyuean, longyongchao, wangshu1, libin, wuxiliang}@yy.com

[§]zzc@jyu.edu.cn

## ABSTRACT

In recent year, the write-heavy applications is more and more prevalent. How to efficiently handle this sort of workload is one of intensive research direction in the field of database system. The overhead caused by write operation is mainly issued by two reasons: 1) the hardware level, i.e., the IO cost caused by logging. We can't remove this cost in short period 2) the dual-copy software architecture and serial replay. The born of log as database architecture is originated to overcome the software defect. But existing systems treating log as database either are built on top of special infrastructure such as infiniband or NVRam (Non-Volatile Random access memory) which is far from widely available or are constructed with the help of other system such as Dynamo which is lack of flexibility. In this paper we build only write-once key-value system called LogStore from scratch to support our instant messenger business. The key features of LogStore include: 1) a single thread per partition executing mode, which eliminates the concurrency overhead; 2) log as database to enable write-once feature and freshness on the standby. We achieve high availability by embedding replication protocol other than dependent on other infrastructure; 3) fine-grained and low overhead data buffer pool management to effectively minimize IO cost. According to our empirical evaluations LogStore has good performance in write operation, recovery and replication.

## 1. INTRODUCTION

As has been observed by Yahoo! [1], in web2.0 era more and more application workload shifts from read-intensive to write-intensive. In the period of 2010 to 2012, the ratio of write operations in their applications have raised from about 10% to 50%. This phenomenon in a sense comes with social network workload, e.g., encountered by Facebook, Twitter and WeChat, wherein different users issue many updates and later each of them requests to retrieve all new posts by a single read operation. Our Messenger BCM Messenger born for security and speed has encountered same problems as above apps and it has great demand on high performance and high availability. Except social network scenario, the infrastructures ingesting stream data (such as user clicks and sensor readings) face large write-intensive workloads. Their most important feature is able to persist all inputs. Lack of strong ingestion power, system has to throw away part of data or block user request, which will hurt the experience of users. Furthermore, some kind of workload which also greatly affects our daily life such as financial transaction or promotion in e-commerce are write-intensive in nature. Therefore, it is attractive that the storage system is featured with high write throughput, low latency, fault-tolerance and low storage usage acquirement while offering instant data recovery capability.

Storage system widely adopts write-ahead logging [2] approach to maximum write throughput while providing durability guarantee. In this architecture, the write operations could be transformed to sequential log file write. It is known as no force policy, under which when the operation(s) is finished, the data modification(s) is no need to be written back, only the incremental log is flushed. In case of system crash, the log can be used to recover the system. This architecture is referred to as dual-copy storage. The dual-copy storage has three disadvantages. Firstly, Although it can defer modified data to disks, all the data have to be flushed into the persistent storage when the time or capacity threshold reaches, i.e., after checkpoint interval timeout or the dirty page percentage exceeds high water mark , which would cause the write bottleneck in write-heavy applications. Secondly, dual-copy storage requires careful orchestration between the log and data store, which increase the complex software logic and unavoidably limits the performance of memory-resident workloads. Besides, the complex software comes with the more code maintenance and testing costs. Actually, as is pointed out by stonebraker, a DBMS really consists of two DBMSs, one managing the data as we know it and the other managing the log [3]. Last but not the least, under the dual-copy storage there exists a significant gap between the database and log. So when encountering system failures, the log need to be replayed against the database to transform it to the latest status, leading to an obvious outage interval. Dual storage also causes big trouble in high availability setting where the hot standby is prepared to take over primary when the primary crashes and serves read request in order to increase the hardware utilization and read throughput. With regard to the dual-copy storage, backup usually lags behind primary because the shipped log must be replayed in the backup, which is the same as when system failures happens.

In order to meet high performance and high availability demand of our messenger, we develop LogStore, a Key-Value system following the design philosophy "log is database". We verify LogStore can achieve good read/write throughput and is good for high availability behavior. High availability behavior is defined as about the following two factors: 1) when the primary crashes how quickly the secondary can take over as primary 2) the secondary can afford what kind of freshness if it serves read-only request. Although more and more DBMSs use the log as the unique data repository they are built on top of Amazon S3 or special hardware (e.g. Non-Volatile Memory, Infiniband). We develop LogStore from scratch on commodity environment to test to what extent we get benefit from log as database. In summary, key features of LogStore are as follows:

- One stone, two birds. We develop a high performance Key-Value data store called LogStore. LogStore adopts the log as database idea to maximum write throughput and improve the freshness on the backup. LogStore is built from scratch on commodity machine environment, and replication module is embedded in it, under which condition we can investigate thoroughly to what extend the log as database brings benefits to us.

- To efficiently support the read operation, we integrate a lightweight buffer pool into LogStore. The buffer pool is detached from the low level storage. Furthermore, Log Store doesn't need to track the record access information like CLOCK, and Least Recently Used (LRU) so on. Instead, it utilizes time-space locality to find victim.

- Like VoltDB, we use single-thread-per-partition executing mode to eliminate concurrency control overhead.

The reminding of this paper is organized as follows. In Section 2, we give an overview of related work on log as database for DBMSs and Key-Value stores. Section 3 presents our system design, how LogStore executes the operation, manages buffer pool. In Section 4, we conduct the experimental evaluation of our approach substantiating the claim that log as database is good for performance and high availability behavior. Section 5 concludes the paper.

## 2. Related Work

LogStore involves a wide area of research space such as Key-Value store, Log as database implementation, hot standby and data cache so on. This chapter summarizes the-state-of-art research progress.

*Key-Value store*: RocksDB [4] is developed by Facebook after forking the code from Google's LevelDB [5,6], and it is designed to be an embedded, high-performance, persistent key-value storage system. Facebook's developers optimize rocksDB in terms of level sizes and compression strategies. Beside Facebook, many other companies such as Yahoo!, Linkin, and MongoDB use RocksDB as basic store infrastructure. However, RocksDB uses WAL + data dual copy architecture. As has been confirmed, this store model is not friendly for disaster recovery and high availability behavior. Faster by Microsoft is a high performance Key-Value store. Latch-free data structure, seamless integration of secondary storage, and in-place update capabilities contribute to Faster's high performance. Faster adopts a novel second chance like buffer managers. When a record is fetched from disk it stays in the hot region called hybrid log, where the record can be updated in place. As time goes by the record is moved to read-only region and finally is evicted. LogStore uses very similar idea for buffer management. LogStore randomly picks a victim to cooling region and LogStore not only focuses on point query but also supports the range query. Bitcask [7] is an append-only Key-Value store and uses log as the only data repository as well, but it must assure that the memory is big enough to hold key space. Bitcask deliver the buffer management to OS. Like Faster [8], Bitcask just supports the point query.

*Log as database*: The development group of Amazon Aurora [9] deem the bottleneck of high throughput data processing is moved from compute and storage to network. So Aurora only writes redo log cross network into Amazon S3 to reduce network IO and deliver logging, recovery to storage node. By means of this practice, Aurora can give promise on service level agreement and do a lot of optimization on the computing engine such as concurrency control, latching. But Amazon Aurora is designed to run on the Amazon Web Service ecosystem, we lack deep insight to what extend the log as database bring benefits to high availability behavior and performance. LogStore is built from scratch, and we demonstrate that log as database design philosophy is beneficial even in the commodity computing environment. LogBase [10] is another DBMS adopting log as database. It is developed on top of HBase. Unlike HBase, LogBase only write data once to optimize write throughput. The main contribution of LogBase are devising an in-memory multi-version index to enable efficient data retrieve and supporting transactional semantics. LogBase must do compaction to support range query, and compaction will greatly affect the system performance. In LogStore, the compaction is done likewise. In the near future, we plan to adaptively do compaction adaptively [19]. Hyder [11] decouples transaction processing from storage and scales its database in shared-flash environment on data-sharing cluster. However, LogStore target at the commodity machine not just flash-base storage. The most related to our work is Query Fresh [12] developed by university of Toronto. Query Fresh uses log as database to construct a high performance standalone system and takes hot standby into consideration, i.e., it provides high performance on the primary while achieving good high availability behavior. Replay in Query Fresh just install the log record in the data array, so the replay cost is negligible. Furthermore, the replay is parallel between threads to accelerate the process. The biggest drawback of Query Fresh is that it is developed under the high-end cluster, where the machines are equipped with Non-Volatile Memory and connected by infiniband. LogStore has no such constraint and aims to be deployed in the commodity environment

*Replication*. In the distributed system replication is one of the most complex techniques. Replication strategies are divided into two categories: active [13] and passive [14,15]. The implementation of active replication is simpler as it specifies network rounds when committing operations and it usually can realize low latency. The biggest disadvantage of active replication is that it requires the operation series must be deterministic. This constraint contradicts the real world workload, so most engineering practice adopt passive replication. LogStore uses raft-like replication protocol.

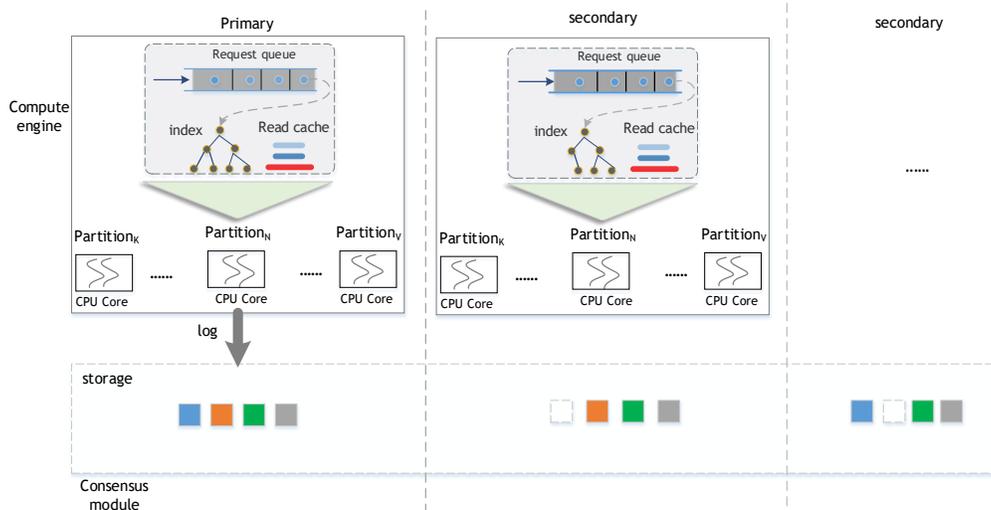

Fig.1. System overview

## 3. System Architecture

LogStore is a concurrent latch-free key-value store that is designed for high performance and good high availability behavior. We use single-thread-per-partition execution model to eliminate synchronization cost, and explore extendedly idea of log as database to develop system with desirable feature. Furthermore, to facilitate the read operation we integrate a lightweight read buffer into the LogStore. To support range query, LogStore does compaction periodically. So the data in the LogStore are separated into two parts, sorted and unsorted. Fig. 1 shows the overall architecture of LogStore. There is a queue for each partition to queuing the incoming request, and LogStore consists of ART index, where each leaf entry holds offset to records and some metadata. To efficiently support read operation, LogStore is also equipped with read buffer pool. Consensus module is embedded in the system for the purpose of high availability.

**Queue.** We use single-thread-per-partition executing model to eliminate synchronization cost. When Read/write request comes, it is put into the request queue. Once former request is finished the system fetches one request from queue and executes the corresponding operations.

**Index**. We adopt the adaptive radix tree (ART) [16] for the purpose to get the needed record at most one IO. ART is an efficient in-memory index and very space saving by adaptively changing the node size according to the node space consumption. Mostly importantly, as is demonstrated by the paper [17], the lookup performance of ART exceeds highly optimized, read-only search trees and it also efficiently supports the deletion and insertion. But in ART, the scan operation is not supported very well, in future work we plan to enhance this feature of ART.

**Storage**. We divide the system into compute engine and storage component. By this means of separation, we can independently optimize the computing and storage. Like Aurora, we only write log into storage. This practice is good for network bandwidth and we can do a lot of optimization on the top of this architecture e.g., the compaction operation in the storage can be moved away to another machine and the computing engine can do optimization against index, read cache and so on independently.

### 3.1 Execution Mode

Multi-threading, elaborate synchronization is the norm for high performance system design. Inside the traditional RDBMS, this norm is used to extremes. Beside this design philosophy, the low level hardware plays an important role. At the time of born of RDBMS, the computer usually featured low speed external device and the main-memory is measured by KBs. The speed gap between disk and main memory is orders of magnitude. The bottleneck for storage system is disk IO. So in order to obtain a high performance system, the key is to run tasks in parallel as many as possible, expecting the multiplexing concurrent tasks to hide the IO latency. The eternal theme in the storage system is how to design efficient buffer pool, fine-grain locking, and sophisticated latching and logging schemes. But now the situation has changed: machine equipped with hundreds of gigabytes is not uncommon. Many types of workload can be done inside the main memory, such as OLTP transaction, and these types of workload especially for Key-Value store only involves point query such as searching or updating few records. These operations itself take only microseconds. Under this case, modules such as locking, latching, buffer management etc. could contribute most of execution time as is verified by work[18] where Harizopoulos et al showed that for typical OLTP workload the real work done by transaction only takes up to 12%, other time consumed by concurrency control, latching , logging and so on. We follow their research finding. LogStore uses single-thread-per-partition execution mode to eliminate synchronization cost. The thread assigned to some partition is bound to specific CPU core to reduce the context switch cost. When a read/write request comes in, it is put into request queue. Sometime later, the system fetches the task and finishes it, during which time the task can't be grabbed and executes on that core exclusively.

3.2 Index

In LogStore, the log files are divided into two parts. One is sorted, inactive segment; the other is unsorted, active segment. LogStore sequentially write the record into active part. LogStore periodically merge the unsorted part with sorted part to remove deleted items and to facilitate range query. The records in both parts are indexed by the ART tree to support accessing requested data efficiently. As is illustrated in the Fig.2, ART tree is built on top of log files, and log files are separated into two parts. The leaf entry is in the form of <*key*, *offset*> tuple where *offset* points the location of corresponding value. In future, we can integrate time information into key to support multi-version mechanism. Read/write flow are as follows.

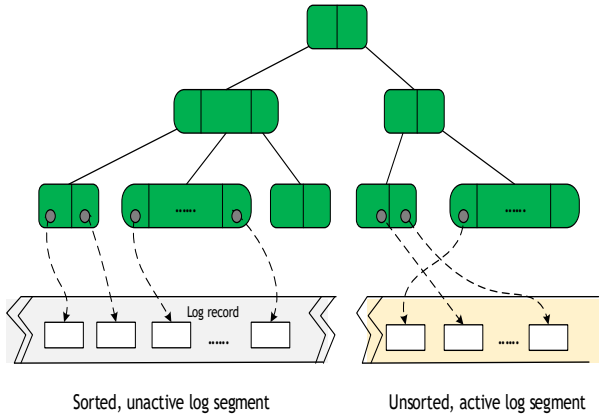

**Fig.2. index for data**

*Write*. When a modification request (Insert, Update) arrives, the request will be put into the corresponding partition queue and corresponding replication channel. Once the thread bound to this partition is idle the request is fetched from the queue. First, the system transforms request into tuple in the format of<*key*, *offset*>. After the log record has been persisted, its starting offset in active log segment are returned so that the LogStore subsequently updates the in-memory index of the corresponding partition. Sometime later, the index servers in retrieve of data records in the log. To reduce the interaction with peripheral, we integrate a lightweight buffer into the LogStore. So, the read operation seldom direct to index file because of buffer. The read flow will describe the detail. Like Amazon Aurora, LogStore never write back the data in the buffer. It just use the buffer to improve read performance. By always replacing older data version in buffer, the application is assured to read latest data.

*Read*. LogStore supports GET (key) semantics, where users must specify the key. To answer the request, LogStore will first check buffer whether the requested value of corresponding key exists in the buffer. If buffer contains the value, it is returned to client. Otherwise, LogStore will turn to index to locate where the value is place in log. Based on the fact that LogStore will replace older version in buffer, the application can get the latest version. In future, LogStore will support operation of getting historical data by using prefix search. In our architecture, the buffer is crux to system performance, so a low cost, high efficient buffer design is very import. We adopt a second-chance like buffer management strategy. Our algorithm doesn't need to maintain access information for record and can reduce the running cost to minimum. In the present of read buffer, LogStore can process most queries in memory. If cache miss happens, the index can assure that the query can be processed by at most one IO.

*Delete*. Like insert and update, the delete operation is also put into request queue. To remove a record in LogStore, two components in LogStore are involved. Firstly, the record should be evicted from buffer so that for the following query, it can't be found in the buffer pool. Next, the index entry of the record must be deleted from ART tree. After these two steps, the query can't find the corresponding data. In order to persist the operation, LogStore should write a record into log file to indicate record with such key is removed.

3.3 Buffer

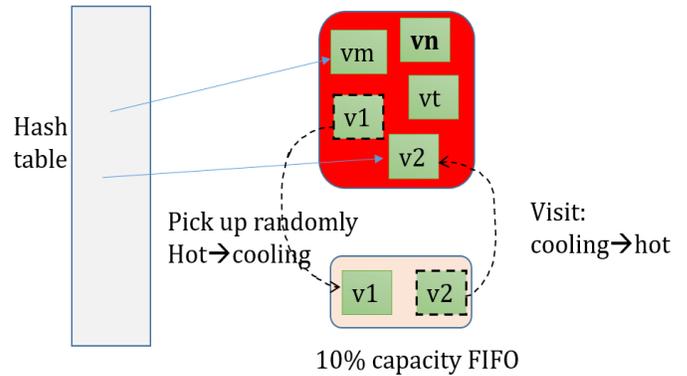

**Fig.3. buffer management**

We maintain an in-memory cache and performance heavily depends on its efficiency. Several caching protocols have been proposed in the context of buffer pool management in databases and virtual memory management in operating systems such as First-In First-Out (FIFO), CLOCK, and Least Recently Used (LRU) Protocol. Except FIFO, All of above algorithms require fine-grained access information statistics in order to work efficiently. LogStore achieves a good cache behavior at a per-record granularity without any such overheads by using a second-chance-like algorithm. The buffer pool is divided into two region. One is called hot region, the other one is cooling region. As illustrated in Fig.3, cooling region take up the 10 percentage capacity and it is organized as FIFO queue. When the buffer pool capacity threshold reaches, the system randomly picks a record and put it in cooling region. If the record isn't touch during some period (from head of queue to tail of queue), we deem this record is cold. So the record can be throw away. The access of the record would make it hot, causing it to be moved to hot region again. By using the space-time locality, the cost of maintaining buffer pool is minimized.

3.4 Replication Module

### *3.4.1 Freshness*

Traditional data + WAL architecture would incur additional replication cost. Let's first describe what would happen if applying replication algorithm on dual-copy system. Taking raft algorithm as example, it is divided into two phases as illustrated in Figure 4:

Although primary can utilize batch and pipeline technique to minimize IO cost secondary faces different problem connecting tightly with dual-copy storage scheme. As each secondary receives log entries sent from primary it will append those log entries into disk. The newly coming log entries can't be applied until they are decided. As depicted in phase 2, the data committed at primary is not immediately available at secondary until it receives the decided point and fully replay those log entries.

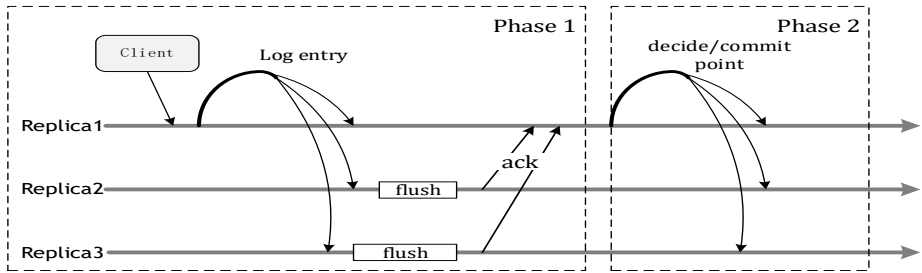

Fig.4. data replication flow

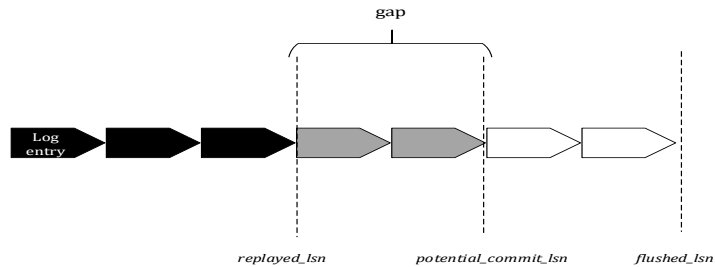

Fig.5. gap between primary and secondary

**Phase 1:** When new requests from client arrives, primary packs a batch of log entries, appends them to disk and replicates to other replicas by calling *AppendEntries* RPCs. After the primary node acknowledges the log entries have been successfully replicated to the majority of replicas, it can commit this transaction and reply to client.

**Phase 2:** The primary firstly update its local commit point. In the next round the primary send commit point by piggyback, then secondaries can calculates the gap between its own commit point and commit point of primary. Now, secondary loads decided log entries from disk (an optimized way is caching the log index in memory, but still pay cost of one more memory copying) and call applicator to replay the log entries against memTable (which often organize data by means of SkipList or B+-tree ). This indicates that the log entries received and flushed in this round must be re-read for applying in the next round, which incurs lags between primary and secondary.

When the read-only requests are directed to secondary nodes, it may read much earlier snapshot. As shown in figure 5, there exists three kinds of log entries. One kind is those replayed log entries reflecting their operation on data store; the second kind is log entries committed on primary, but not yet replay, which we refer it as potential committed log; another kind is log entries that have not been decided. We use three variables to define boundary of those three kinds of log entries. (1) *replayed_lsn*: log entries before this point have been applied; (2) *potential_commit_lsn*: log entries between *replayed_lsn* and *potential_commit_lsn* have been committed on primary, but the commit notification has not been piggybacked to secondary. (3) *flushed_lsn*: log entries from *potential_commit_lsn* to this point have not been decided. Read request with lsn $l$ got from primary will face the following three cases:

Case 1: $l < replayed\_lsn$. In this case, read request can read what it needs and return the result quickly back to client.

Case 2: *replayed_lsn* <= *l* < *potential_commit_lsn*. If lsn of read request falls into this interval, the read request will be blocked until the follower receives commit decision and log entries in this interval are all fully replayed.

Case 3: *l* >= *potential_commit_lsn* or *l* >= f*lushed_lsn*. When lsn of read request exceeds *potential_commit_lsn*, read request may wait a long time to get commit decision or acquires non-existence data (*l* >= *flushed_lsn*). In both cases, the request will be discarded.

But it is a different state in Case 2 when treating log as database. The system only needs to advance *replayed_lsn* to *potential_commit_lsn*, representing that log before this LSN have been applied and can serve the query. The gap as illustrated in figure 5 is almost eliminated between the primary and secondary. Imaging this situation when a read request obtain latest read view lsn *l* from meta server, it is then directed to secondary where it can just compares *potential_commit_lsn* with *l*. if *l* is no great than *potential_commit_lsn* read request can utilize the in-memory index to retrieve demanding data other than blocked to wait. The log as database design improves freshness.

### *3.4.2 Replication technique*

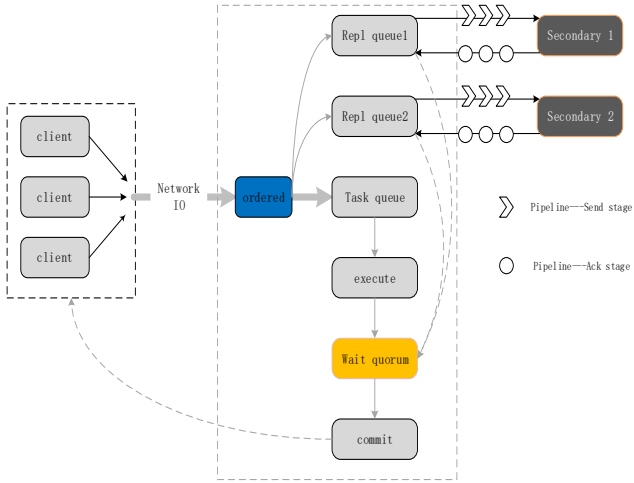

**Fig.6. replication overview**

In LogStore, raft-like protocol is adopted for data replication. A core design tenet for our replication module is to maximum the system throughput and minimize latency of write request. The most two import techniques adopted in LogStore are batching and staged stream pipelining. Since modern database servers are often machines with powerful processing ability pipelining can improve the resource utilization. In order to reduce the average per-request overhead, the primary of partition packs several requests in one task. The flow of data replication in LogStore is as follows. (1) The responsibility of primary server is to order request, execute local commit and await the acknowledgement from secondaries. LogStore will replicate the request to secondaries as soon as it receive the request from client. This step is done in parallel with local commit to avoid serialization waiting in local commit and acknowledgement from other storage nodes. Note that the storage node for primary and secondary can be deployed different from processing node. (2) In our initial version, data can't be sent before the prior ack is received. Under this implementation, the latency of each operation is twice than staged stream pipelining in theory. We then separate the replication into two stages, i.e., data sending and ack receiving, so they can run simultaneously. (3) When the received acks reaches majority system returns the result to client. Furthermore, we optimize LogStore replication module from some perspectives of engineering, e.g., the staged stream pipelining utilizes efficient gather IO to reduce IO system calls and counting quorum component uses lock-free data structure and so on. When serving read-only request, the replica in LogStore can provide desirable freshness duo to log as database design. As above stated, LogStore doesn't need to replay the log to produce "real" data. The request can query ART tree to read the latest log file to get the value. Our replication algorithm is depicted detailedly in figure 7. For the convenience of expressing algorithm, we give the related notation in the following.

$\mathcal{M}_p$: modification requests set maintained by *p*th partition.

$u_p^i$ : *i*th modifying operation (including insert, update and delete) sent to *p*th partition

$\mathcal{R}_p$ : replication set for *p*th partition

$\mathcal{C}_p$, $\mathcal{C}_{p\_repl}$, $\mathcal{C}_{p\_ack}$, $\mathcal{C}_{p\_r}$ : completed tasks for modifying, replication, waiting ack and reply set respectively.

$u_p$ : a batch of modification operation for *p*th partition

$state_p^{i-k+1} \leftarrow applicator(u, state_p^{i-k})$ : the state transits from $state_p^{i-k}$ to $state_p^{i-k+1}$ under the effect of applying modification operation *u* by function *applicator*.

$\rho(u)$ : pack a batch of modification requests and send to other storage node by one system call.

$\mathcal{A}_p^u$ : modification *u*'s ack set for *p*th partition

$\mathbb{A}_p$: set of all modification operation *u*s' ack set

LogStore consists of four servers: tcpServer, replicationServer, replyServer, stateMachServer. When a request is issued from client tcpServer checks the request and if passed, it accepts the request and puts it into corresponding task set $\mathcal{M}_p$ and replication set $\mathcal{R}_p$ (RequestDispatch, line 3-4).stateMachServer will get a batch of task from task set (ProcessTask, line 4) and use the function applicator to change the state machine temporarily (ProcessTask, line 7). This temporary state will be persisted when the task is successfully replicated over a majority of group members. In order to indicate whether the transition can be replied to client, i.e., whether this state can be persisted or not, the stateMachServer also wraps the modification *u* as reply object $\tau_p^u$ and add it into *p*th partition's reply set $\Upsilon_p$ (ProcessTask, line 8).

**Replication Algorithm: replicate log record to other storage nodes**

**Input:** file descriptor fd
1. **Procedure** RequestDispatch(fd) /*dispatch a task to certain channel*/
2.     **while** LogStore.IsRunning **do**
3.        $M_p \leftarrow M_p \cup \{u_p^i\}$ /*receive ith request for pth partition*/
4.        $R_p \leftarrow R_p \cup \{u_p^i\}$ /*put ith request into pth channel*/
5.        increment LSN of replication channel

**Input:** set of pth partition requests $M_p$
1. **Procedure** ProcessTask($M_p$) /*process one batch of task*/
2.     **while** $M_p - C_p \neq \emptyset$ **do**
3.        $U_p \leftarrow \emptyset$
4.        $U_p \leftarrow$ pick a batch of requests from $M_p - C_p$
5.        $C_p \leftarrow C_p \cup U_p$
6.        **foreach** $u$ in $U_p$ **do** /*processing a batch size k*/
7.           $state_p^{i-k+1} \leftarrow applicator(u, state_p^{i-k})$
8.           $r_p \leftarrow r_p \cup \{\tau_p^u\}$ /* add u's reply obj $\tau_p^u$ into waiting reply set $r_p$ */
9.           add ack into u's ack set $\mathcal{A}_p^u$ in pth replication channel /*local ack*/

**Input:** set of pth replication requests $R_p$
1. **Procedure** replicateSendStage ($R_p$)
2.     **while** $R_p - C_{p\_repl} \neq \emptyset$ **do**
3.        $U_p \leftarrow$ pick a batch of requests from $\{R_p - C_{p\_repl}\}$
4.        $C_{p\_repl} \leftarrow C_{p\_repl} \cup U_p$
4.        send $\rho(U_p)$ /*pack modification messages and send by one sys call*/
5.        **foreach** $u$ in $U_p$ **do**
6.           $A_p \leftarrow A_p \cup \mathcal{A}_p^u$ /* add u's ack set $\mathcal{A}_p^u$ into waiting ack set $A_p$*/

**Input:** set of pth partition acknowledge $A_p$
1. **Procedure** ReplicationRecvAndAckStage ($A_p$)
2.     **while** $A_p - C_{p\_ack} \neq \emptyset$ **do**
3.        $A_p^i \leftarrow$ pick a batch of waiting ack from $\{A_p - C_{p\_ack}\}$
4.        $C_{p\_ack} \leftarrow C_{p\_ack} \cup A_p^i$
5.        **foreach** $\mathcal{A}_p^u$ in $A_p^i$ **do**
6.           add ack into $\mathcal{A}_p^u$ /*receive an ack for u's ack set $\mathcal{A}_p^u$ */

**Input:** set of pth partition reply $r_p$
1. **Procedure** ReplyProcess ($r_p$)
2.     **while** $r_p - C_{p\_r} \neq \emptyset$ **do**
3.        $\tau_p \leftarrow$ pick a batch of waiting reply from $r_p - C_{p\_r}$
4.        $C_{p\_r} \leftarrow \tau_p \cup C_{p\_r}$
5.        **foreach** u's reply obj $\tau_p^u$ in $\tau_p$ **do**
6.           fetch u's ack set $\mathcal{A}_p^u$ by $\tau_p^u$
7.           **if** $|\mathcal{A}_p^u| \geq N/2 + 1$ **do**
8.              send reply to client

Fig.7. replication algorithm

The last step in procedure ProcessTask is adding local ack into u's ack set $\mathcal{A}_p^u$ belonging to pth partition's replication channel. As above stated, we adopt stage stream pipeline technique in replication where the replication request (procedure ReplicateSendStage) and receiving of replicating successfully ack (procedure ReplicateRecvAndAckStage) can run in parallel. ReplicationServer picks a batch of tasks from replication request set $\mathcal{R}_p$ (ReplicateSendStage, line 3) and send those requests within one system call by using gathering IO technique (ReplicateSendStage, line 4). Then, put task u's replication ack set $\mathcal{A}_p^u$ into waiting ack set for all tasks. (ReplicateSendStage, line 6). At the same time, replicationServer receives ack by means of ReplicationRecvAndAckStage procedure which add ack into ack set $\mathcal{A}_p^u$ for corresponding modification request u (ReplicationRecvAndAckStage line 6). In replyServer, each reply object $\tau_p^u$ produced in processTask procedure will be checked (procedure replyProcess line 5). Firstly, replyProcess fetches the u's ack set $\mathcal{A}_p^u$ through $\tau_p^u$. If the cardinality of $\mathcal{A}_p^u$ LogStore can send reply to client indicating that the modification request is finished (procedure replyProcess line 7-8).

## 4. Experiment Evaluation

In this section we will empirically investigate LogStore in terms of throughput, scalability, replication and recovery.

- First, we compare overall throughput and multi-thread scalability of LogStore with LogBase and HBase by varying the memory consumption.
- Second, high availability behavior of LogStore is examined. We check to what extend secondary lags behind the primary. If using secondary serves read-only request, what kind of freshness can secondaries afford?
- Third, we evaluate recovery time of LogStore to survey the benefit that log as database brings to us.

4.1 Experiment Setup

Experiments were performed on Alibaba Cloud cluster including 3 machines, each with 16 core processor, 32 GB of physical memory, 600 GB of SSD capacity connected through a 10 Gigabits Ethernet by default. But due to the bandwidth permitted by Alibaba Cloud, we can use only the 3 Gigabits network bandwidth. Source code of LogBase (developed by NUS database group team) downloaded from github can't run properly, we fix those bugs. In all, the problems are as follows:

(1) mismatch in the HDFS and client of LogBase. Originally, HDFS (v 0.20.2) used by LogBase is incompatible with client. After tries, we replace the client and HDFS with version 0.20.205.

(2) wrong implementation of function CreateFileNum where the code piece

**int ret = (int) (System.currentTimeMillis() / 100**

will result in overflow and is wrong in logic.

(3) every operation has to consult the meta, causing performance issue. We add cache into client to bypass this problem.
(4) `get` (byte [] key) API will iterate all keys greater than key.
(5) LogBase will write its log into hdfs, and write HLog (WAL for HBase) one more time, i.e., write log twice. We modify it to write log once.

The modified LogBase is available at github https://github.com/logkv/logbase_changed.

LogStore is implemented using C++ in 10K lines of codes, adopting the idea log as database. For in-memory index, we use ART Tree and all access will check buffer first, if not found, then data position is determined through index. Data replication function is embedded into LogStore, so we can evaluate log as database design in all aspects. HBase (version 2.1.0) and HDFS (version 3.1.1) is used in the experiment. The block cache consumed by HBase is 8GB. For HDFS, all settings are kept default, especially the chunk size is 64MB and the replication factor is 3. The data used in experiment is produced by YCSB. Size of data record is about 1KB, and range of key is $(0, 2 * 10^9)$. We warm up system before each experiment.

4.2 Benchmark

### 4.2.1 Write Performance

Figure 9 left compares the performance of LogBase, LogStore and HBase under the different concurrent threads in standalone mode, in which situation we can exclude unrelated factors to study the LogStore internal. LogStore outperformances LogBase and HBase by about 4x under the concurrency of 128 threads. Furthermore, as figure 9 left illustrates LogStore scales line nearly. This is mainly because of one-thread-per-partition execution model, which eliminates synchronization cost and fully utilizes computation resource. Figure 9 right gives the performance of three systems when replication is on. As QPS increases (duo to the more concurrent threads) the latency of LogBase and HBase increase sharply. Peak performance of HBase and LogBase is 32,000 and 37,201 respectively, whereas LogStore's QPS can reach 80,000. Data replication has almost no effect on LogBase's performance. This is duo to asynchronous commit in LogBase, i.e., LogBase can return the result back to client before the data is safely replicated on other storage nodes. On other hands, LogStore and HBase the declination of throughput in different degree. Throughput of HBase drops about 12%, whereas when replication is on throughput of LogStore is half of that of standalone LogStore. Actually, in replication situation latency of HBase is twice more than that of LogStore. LogStore strike a balance between latency and throughput. We also test how dual-copy influences the write performance in figure 10. The memstore of HBase is set to 32MB, 64MB and 128MB. When memstore is full, HBase has to flush the memstore into persistent storage, which incurs more write overhead. The write performance of HBase is greatly affected in case of using 32MB memstore. Under this setting, write performance of HBase is only half compared to that using 128MB memstore. LogStore and LogBase only write the data once, so they don't have that problem.

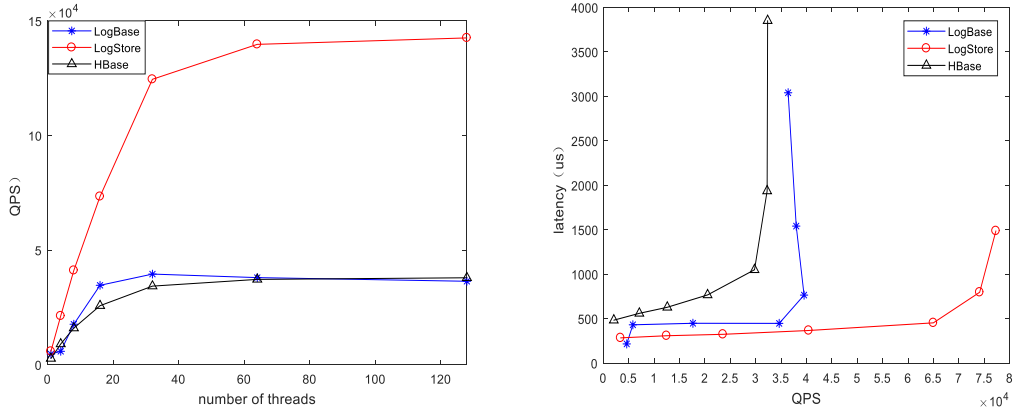

Fig.9. write performance

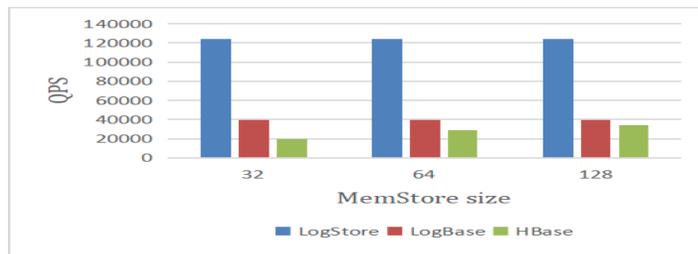

Fig.10. Memstore's effect on HBase

### 4.2.2 Buffer Pool Behavior

We perform a study to compare FIFO, LRU caching protocols with caching behavior of LogStore (TwoStage). The ratio of cache size to data size varies from 0.1 to 0.5. Cache size for key is constant in every experiment. A cached key would be evicted whenever an operation encounters cache miss. For TwoStage when a key is visited, its memory location is recorded in hash table. No more work is needed. Unlike LRU, which should maintain access time and position in LRU chain, TwoStage is simple. When eviction happens, TwoStage randomly pick a key and put it into FIFO queue. We compare TwoStage with LRU and FIFO with respect to cache hit ratio in different cache size. As illustrated in figure 11, TwoStage is better than two other popular cache management protocol for two access patterns: uniform, zipfian in terms of hit ratio. Besides, TwoStage is light weight because it doesn't need maintain access statistics.We deem this feature is good for overall system performance.

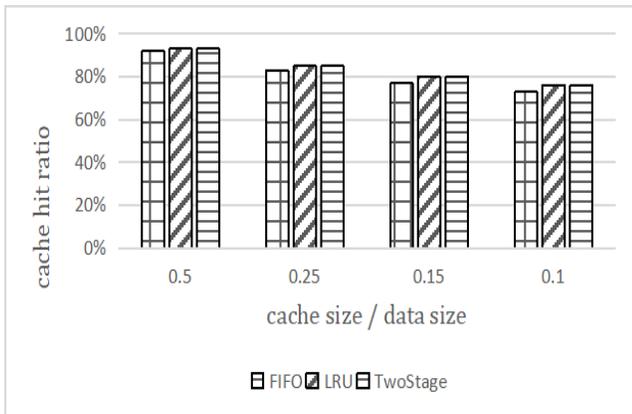

Fig.11. buffer pool performance

### 4.2.3 Recovery

*Freshness*. This section surveys replication feature. We focus on gap between the leader and follower. The freshness can be measured by freshness score like[12] which is defined as follows:

$$freshness\ score = bLSN / pLSN$$

where bLSN refers to the LSN on follower(backup) and pLSN refers to the LSN on leader(primary). Wang et al. synchronize clock on each node and report read view LSN every 20ms. For example, at some time t, leader's LSN is 1,000 and follower's LSN is 800, the freshness score is 0.8. Likewise, we probe time-LSN pair and fit a line to them using matlab's cftool. So, we can measure to what extend secondaries lag behind primary.The reason that we use LSN instead of committed LSN or replayed LSN on follower is log as database design. Think real work environment where client get read view LSN from meta sever, then read from secondary. Maximum LSN of record in log file in secondaries can be returned back immediately without replay. So, we think the LSN on leader and follower is more suitable for log as database design to compute freshness degree. We plot 8 pairs on primary, fit a line to them and select pairs of same time on secondaries. As is shown in figure 12, the lags between primary and secondaries is negligible, the data committed on primary can be read immediately on the secondaries. If primary crashes, the secondary take over control and serves client quickly as well.

*Recovery time.* Now, we study the recovery metrics for LogBase, LogStore. The snapshot cost (write the in-memory index into durable storage) and recovery time should be measured to investigate system recovery feature. But for log as database design, the snapshot cost is proportional to recovery time, so we only need to give recovery time. For LogBase and LogStore, the recovery cost is strongly related to the amount of log records to read because they don't need to apply log record against data store to produce latest data file. We utilize the fork system call to produce snapshot for LogStore. As figure 13 depicted, LogStore is desirable in recovery performance. The cost to reload the snapshot and recovery time of LogBase is nearly 10x more than LogStore. Therefore, the checkpoint cost by LogBase is 10x more than that of LogStore. When the data size exceeds 64G, LogBase takes one more hours to produce snapshot. So, we don't measure the time for LogBase in data size greater than 64GB cases.

Above experiments are conducted above single partition. We can deem if recovery is done in parallel among partitions LogStore can handle recovery for TB data size scale within 1 minute

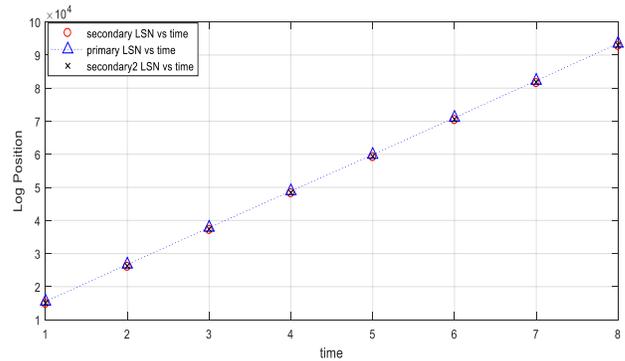

Fig.12. freshness gap

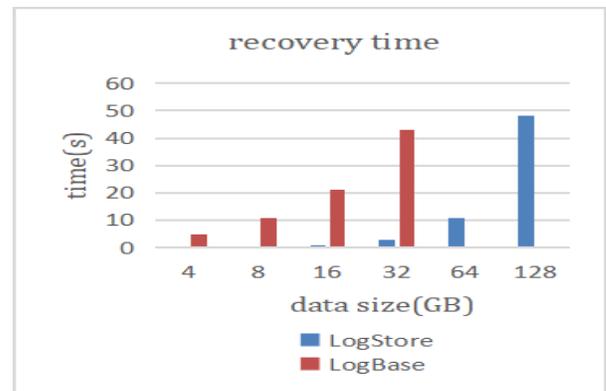

Fig.13. recovery time

*4.2.4 Read*

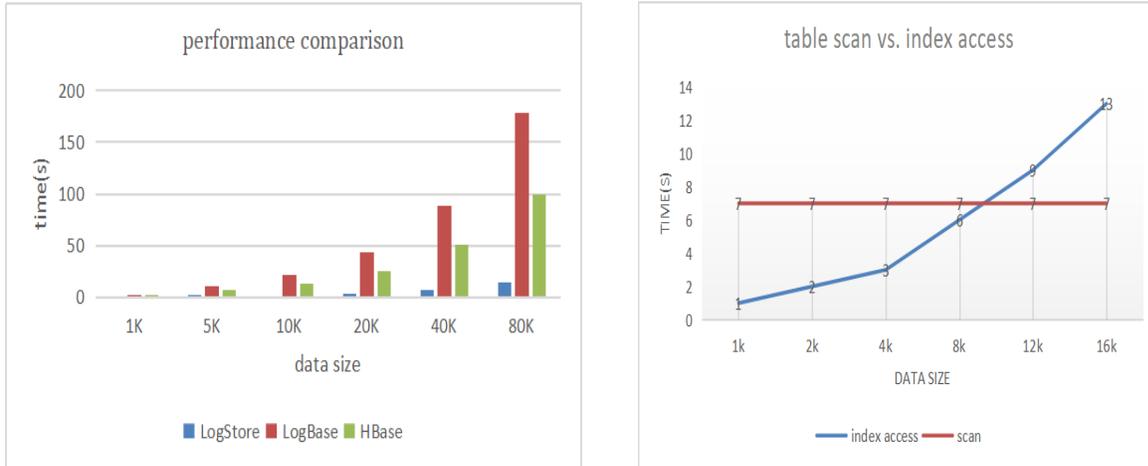

Fig.14.random read

To verify the index effect, we investigate the performance of random access for LogStore, LogBase and HBase. We vary the batch size from 1K to 80K. The total time by three systems is composed of network latency and execution time. We render YCSB to return result to client once a random access request, i.e., the number of acquired records is equal to the network round that client interact with server. LogStore and LogBase can utilize in-memory index to locate precisely position of record. On the other hand, HBase should scan the block and use bloom-filter to get desirable data. From Figure 14 left, we can see that the performance of HBase is better than LogBase because data volume is not enough to incur multiple data level and bloom filter can greatly reduce the data block scan. LogStore is 10x better than both duo to simplicity and high efficiency in execution path. Actually, the LogStore is approximate to the ideal performance.

In this case the TPS is 5698 observed in the experiment, but network latency is 156us, and the ideal performance is about 6400.Then, we modify the YCSB benchmark, allowing it to randomly get a batch of record. In this experiment, we want to probe the cost of index access. The batch size of index access ranges from 1K to 16K. When the data size acquired is smaller than 8K, the index access is more advantageous. But when the data size exceeds 8K, the sequential access is superior to index access. We also integrate this simple heuristic into LogStore, in which when the access data portion is greater than 1/10, the table scan method is launched.

## 5. Conclusion & Future work

This paper presented a single storage design for key-value storage systems called LogStore, which was developed aiming to support our instant messenger. The design follows log as database philosophy, in which all persistent data is stored only in indexed log file, unifying data and log in a general approach. This novel storage architecture decouples in-memory data structures from their persistent representation, eliminating many of the overheads associated with traditional WAL + Data systems design. The log-structured approach also greatly improves recovery capabilities in comparison with tradition three-phases designs, i.e., scan, redo, undo. Because the indexed log contains precise information required to retrieve data items in their most recent, consistent state, recovery from a system failure just rebuild the index. Furthermore, unlike most state-of-the-art approaches, there are no need to do checkpoints, write dirty data back and force WAL.

By merging state-of-art data management techniques such as fast in-memory access methods, log-structure data organization, and buffer management in a novel way, LogStore achieves a moderate spot between modern in-memory database systems and traditional disk-based approaches with respect to design and performance. Under this design, LogStore can reach tens of millions of consistent read/write operations per second. Meanwhile, read/write latency in LogStore is limited to less than 20ms. Furthermore, its high availability behavior is desirable, i.e., freshness gap between primary and secondary is negligible. Our future work is to elaborate the optimizer and let the storage do compaction according to query adaptively.